# Mitigating of Modal Instabilities in Linearly-Polarized Fiber Amplifiers by Shifting Pump Wavelength

Rumao Tao, Pengfei Ma, Xiaolin Wang, Pu Zhou, Zejin Liu

*Abstract*—We investigated the effct of pump wavelength on the modal instabilities (MI) in high power linearly-polarized Yb-doped fiber amplifiers. We built a novel semi-analytical model to determine the frequency coupling characteristics and power threshold of MI, which indicates promising MI suppression through pumping at an appropriate wavelength. By pumping at 915nm, the threshold can be enhanced by a factor of 2.36 as compared to that pumped at 976nm. Based on a high power linearly-polarized fiber amplifier platform, we studied the influence of pump wavelength experimentally. The threshold has been increased by a factor of 2 at 915nm, which agrees with the theoretical calculation and verified our theoretical model. Furthermore, we show that MI suppression by detuning the pump wavelength is weakened for fiber with large core-to-cladding ratio.

## I. INTRODUCTION

HIGH power linearly-polarized fiber sources are required in many applications, such as coherent lidar system, nonlinear frequency conversion, coherent beam combining architectures [1-4]. Linearly-polarized fiber laser with 1kW output power has been achieved recently [5], which employed a pairs of high power monolithic fiber Bragg grating (FBG) in a Fabry-Pérot cavity. It is a technological challenge to design FBG that can withstand multi-kilowatt power, and further power scaling through oscillators may confront with some technological difficulties. Fiber laser systems based on MOPAs are typically capable of reaching high output powers, and most of the high power fiber laser systems with random polarized output are based on MOPA at the moment, which has achieved output power as high as tens of kilowatt. However, power scaling of linearly-polarized MOPAs to multi-kilowatt level is currently limited by the onset of mode instabilities (MI) [6], which deteriorates the pointing stability and the beam quality significantly. Although lot of work has been carried out to deal with MI experimentally and theoretically [7-15] after the first mention of this phenomenon, few methods suitable for mitigating MI effectively in monolithic fiber amplifier with standard step-index large mode area (LMA) fiber have been proposed [16].

Recently, experiments shows that MI threshold power in monolithic fiber amplifier can be increased by detuning the pump wavelength [6, 11]. To the best of our knowledge, no detailed theoretical work has been carried out. In this paper, based on a novel semi-analytical model and a monolithic polarization-maintained (PM) experimental platform, the effect of pump wavelength on MI in high power fiber amplifiers has been studied theoretically and experimentally.

## II. THEORETICAL MODEL

In high power fiber laser systems, most of the fibers are weakly guided fibers, where optical fields can be well approximated by linearly polarized (LP) modes. For linearly-polarized fiber lasers, the optical field of the signal propagating in the fiber is expressed in the conventional LP mode representations

$$E(r,\phi,z,t) = \sum_{m=0}^{\infty}\sum_{n=1}^{\infty} A_{mn}(z,t)\psi_{mn}(r,\phi)e^{j(\beta_{mn}z-\omega_{mn}t)} + c.c. \quad (1)$$

where $m$ and $n$ is azimuthal and radial mode numbers respectively. $A_{mn}(z,t)$, $\beta_{mn}$, and $\psi_{mn}(r,\phi)$ are slowly varying mode amplitudes, propagation constants, and normalized mode profiles of $LP_{mn}$ mode. Assuming the case that the fiber amplifiers are operating below or near the MI threshold, we therefore include only the fundamental mode ($LP_{01}$) and one of the two degenerate $LP_{11}$ modes and the subscripts of 01 and 11 are replaced with 1 and 2 for $LP_{01}$ mode and $LP_{11}$ mode, respectively. Then the signal intensity $I_s$ can be written as

$$I_s(r,\phi,z,t) = 2n_0\varepsilon_0 cE(r,\phi,z,t)E(r,\phi,z,t)^* \quad (2)$$
$$= I_0 + \tilde{I}$$

with

$$I_0 = I_{11}(z,t)\psi_1(r,\phi)\psi_1(r,\phi) + I_{22}(z,t)\psi_2(r,\phi)\psi_2(r,\phi) \quad (3a)$$

$$\tilde{I} = \quad (3b)$$

$$I_{12}(z,t)\psi_1(r,\phi)\psi_2(r,\phi)e^{j(qz-\Omega t)} + I_{21}(z,t)\psi_1(r,\phi)\psi_2(r,\phi)e^{-j(qz-\Omega t)}$$

$$I_{kl}(z,t) = 4n_0\varepsilon_0 cA_k(z,t)A_l^*(z,t) \quad (3c)$$

$$q = \beta_1 - \beta_2, \ \Omega = \omega_1 - \omega_2 \quad (3d)$$

The temperature distribution is governed by heat transportation equation, which is given by

Manuscript received December 1, 2014. This work was supported in part by the National Science Foundation of China under grant No. 61322505, the program for New Century Excellent Talents in University.

Rumao Tao, Pengfei Ma, Xiaolin Wang, Pu Zhou, Zejin Liu are all with College of Optoelectric Science and Engineering, National University of Defense Technology, Changsha, Hunan 410073, China (e-mail: chinawxllin@163.com; zhoupu203@ 163.com).

$$\nabla^2 T(r,\phi,z,t) + \frac{Q(r,\phi,z,t)}{\kappa} = \frac{1}{\alpha}\frac{\partial T(r,\phi,z,t)}{\partial t} \tag{4}$$

where $\alpha = \kappa/\rho C$, $\rho$ is the density, $C$ is the specific heat capacity, and $\kappa$ is the thermal conductivity. Since the heat in high power fiber amplifiers is mainly generated from the quantum defect and absorption, the volume heat-generation density $Q$ can be approximately expressed as

$$Q(r,\phi,z,t) \cong g(r,\phi,z,t)\left(\frac{v_p - v_s}{v_s}\right) I_s(r,\phi,z,t) \tag{5}$$

and $g(r,\phi,z,t)$ is the gain of the amplifier

$$g(r,\phi,z,t) = \left[(\sigma_s^a + \sigma_s^e) n_u(r,\phi,z,t) - \sigma_s^a\right] N_{Yb}(r,\phi) \tag{7a}$$

where $v_{p(s)}$ is the optical frequencies, $\sigma_s^a$ and $\sigma_s^e$ are the signal absorption and emission cross sections, $\sigma_p^a$ and $\sigma_p^e$ are the pump absorption and emission cross sections, $N_{Yb}(r,\phi)$ is the doping profile, the population inversion $n_u$ is given in [13].

Assume that the fiber is water cooled, the appropriate boundary condition for the heat equation at the fiber surface is

$$\kappa \frac{\partial T}{\partial r} + h_q T = 0 \tag{6}$$

where $h_q$ is the convection coefficient for the cooling fluid. By adopting the integral-transform technique to separate variables in the cylindrical system [17], (4), combined with (5) and (6), can be solved as

$$T(r,\phi,z,t) = \frac{1}{\pi}\frac{\alpha n_2}{\eta} \sum_v \sum_{m=1}^{\infty} \frac{R_v(\delta_m,r)}{N(\delta_m)} \int_{t'=0}^{t} e^{-\alpha \delta_m^2(t-t')} dt' \tag{7}$$

$$\times \begin{bmatrix} B_{11}(\phi,z)I_{11}(z,t') + B_{22}(\phi,z)I_{22}(z,t') \\ + B_{12}(\phi,z)I_{12}(z,t')e^{j(qz-\Omega t')} + B_{12}(\phi,z)I_{12}^*(z,t')e^{-j(qz-\Omega t')} \end{bmatrix}$$

with

$$B_{kl}(\phi,z) = \begin{cases} \int_0^{2\pi} d\phi' \int_0^R g_0 R_v(\delta_m,r') \cos v(\phi-\phi') \frac{\psi_k(r',\phi')\psi_k(r',\phi')}{1 + I_0/I_{saturation}} dr', & k = l \\ \int_0^{2\pi} d\phi' \int_0^R g_0 R_v(\delta_m,r') \cos v(\phi-\phi') \frac{\psi_k(r',\phi')\psi_l(r',\phi')}{(1 + I_0/I_{saturation})^2} dr', & k \neq l \end{cases} \tag{8a}$$

$$N(\delta_m) = \int_0^R r R_v^2(\delta_m,r) dr, \quad n_2 = \eta(v_p - v_s)/\kappa v_s \tag{8b}$$

where $v = 0, 1, 2, 3…$ and replace $\pi$ by $2\pi$ for $v=0$, $\eta$ is the thermal-optic coefficient, $R$ is the radius of the inner cladding, $g_0$ is the small signal gain and $I_{saturation}$ is the saturation intensity. $R_v(\delta_m,r)$ is given by $R_v(\delta_m,r) = J_v(\delta_m r)$ and $\delta_m$ is the positive roots of $\delta_m J_v'(\delta_m R) + J_v(\delta_m R) h_q/\kappa = 0$. Considering effective refractive index of gain from amplifier, the total refractive index, which attributes to gain ($n_g \leq n_0$) and nonlinearity ($n_{NL} \leq n_0$), can be expressed as

$$n^2 = (n_0 + n_g + n_{NL})^2 \cong n_0^2 - j\frac{g(r,\phi,z,t)n_0}{k_0} + 2n_0 n_{NL} \tag{9}$$

where $n_{NL}$ is given by

$$n_{NL}(r,\phi,z,t) = \eta T(r,\phi,z,t) \tag{10}$$
$$= h_{11}(r,\phi,z,t) + h_{22}(r,\phi,z,t)$$
$$+ h_{12}(r,\phi,z,t)e^{jqz} + h_{21}(r,\phi,z,t)e^{-jqz}$$

with

$$h_{kl}(r,\phi,z,t) = \begin{cases} \frac{\alpha n_2}{\pi} \sum_v \sum_{m=1}^{\infty} \frac{R_v(\delta_m,r)}{N(\delta_m)} \int_0^t B_{kk}(\phi,z) I_{kk}(z,t') e^{-\alpha \delta_m^2(t-t')} dt', & k = l \\ \frac{\alpha n_2}{\pi} \sum_v \sum_{m=1}^{\infty} \frac{R_v(\delta_m,r)}{N(\delta_m)} \int_0^t B_{kl}(\phi,z) I_{kl}(z,t') e^{-\alpha \delta_m^2(t-t') - j\Omega t'} dt', & k \neq l \end{cases} \tag{11}$$

Inserting (1) and (10) into the wave equation, after very tedious but straightforward derivations, we can obtained the coupled-mode equations

$$\frac{\partial |A_2|^2}{\partial z} = \left[\iint g(r,\phi,z)\psi_2\psi_2 r dr d\phi + |A_1|^2 \chi(\Omega,t)\right]|A_2|^2 \tag{12}$$

with

$$\chi(\Omega) = 2\frac{n_0 \omega_2^2}{c^2 \beta_2} \text{Im}\left(4n_0 \varepsilon_0 c \iint \bar{h}_{12}\psi_1\psi_2 r dr d\phi\right) \tag{13a}$$

$$\bar{h}_{kl}(r,\phi,z) = \frac{\alpha n_2}{\pi} \sum_v \sum_{m=1}^{\infty} \frac{R_v(\delta_m,r)}{N(\delta_m)} \frac{B_{kl}(\phi,z)}{\alpha \delta_m^2 - j\Omega} \tag{13b}$$

For the case that MI is seeded by intensity noise, we can obtain

$$\xi(L) \approx$$
$$\xi_0 \exp\left[\int_0^L dz \iint g(r,\phi,z)(\psi_2\psi_2 - \psi_1\psi_1) r dr d\phi\right] \tag{14}$$
$$+ \frac{\xi_0}{4}\sqrt{\frac{2\pi}{\int_0^L P_1(z)|\chi''(\Omega_0)| dz}} R_N(\Omega_0)$$
$$\times \exp\left[\int_0^L dz \iint g(r,\phi,z)(\psi_2\psi_2 - \psi_1\psi_1) r dr d\phi + \int_0^L P_1(z)\chi(\Omega_0) dz\right]$$

where $R_N(\Omega)$ is the relative intensity noise (RIN) of the input signal, $\xi_0$ is the initial HOM content.

### III. NUMERICAL SIMULATIONS

We have calculated the MI threshold power under different pump wavelength. As shown in [18], MI threshold is independent of rear earth dope concentration. To facilitate fast computation and save time, the length of the fiber was taken to be as short as 1m and the dope concentrations of the ytterbium ions ($N_{Yb}$) are adjusted to the minimum value necessary to achieve high efficiency, defined as pump absorption > 0.95. Other parameters used in our numerical simulation are listed in Table I.

TABLE I
PARAMETERS OF TEST AMPLIFIER

| | |
|---|---|
| $R_{core}$ | 10.5μm, 13μm, 15μm |
| $R$ | 200μm |
| NA | 0.064 |
| $\lambda_s$ | 1064nm |
| $h_q$ | 5000 W/(m$^2$K) |
| $\eta$ | $1.2 \times 10^{-5}$ K$^{-1}$ |
| $\kappa$ | 1.38 W/(Km) |
| $\rho C$ | $1.54 \times 10^6$ J/(Km$^3$) |
| $\xi_0$ | 0.01 |
| $R_N(\Omega)$ | $1 \times 10^{-10}$ |
| $\sigma_p^a$ @915nm | $6.04 \times 10^{-25}$ m$^2$ |
| $\sigma_p^a$ @970nm | $4.47 \times 10^{-25}$ m$^2$ |
| $\sigma_p^a$ @973nm | $1.69 \times 10^{-24}$ m$^2$ |
| $\sigma_p^a$ @976nm | $2.47 \times 10^{-24}$ m$^2$ |
| $\sigma_p^a$ @985nm | $3.01 \times 10^{-25}$ m$^2$ |

Fraction of HOM as a function of pump power is shown in Fig. 1. From Fig. 1, we can see that MI thresholds are dependent on pump wavelength. The key results are listed in Table II. Fiber amplifiers pumped at 915nm, 970nm and 985nm increase the threshold by a factor of 2.3, 3 and 3.6 compared to those pumped at 976nm, which agrees with the previous experimental report [6]. This is due to that if the pump light is detuned from the absorption peak at 976 nm, the reduced pump absorption coefficient will lead to lower upper state populations, which tends to increase hole burning and increase the MI threshold power. The suppression at 973nm is not so significant, which is due to the relatively large pump absorption coefficient.

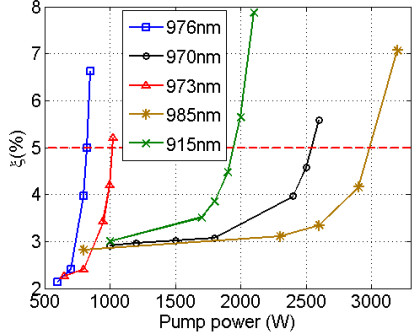

Fig. 1 HOM contents as a function of pump power

TABLE II
MI IN PM 21/400 FIBER

| Pump wavelength (nm) | Coupling frequency (kHz) | Threshold power (W) | |
|---|---|---|---|
| 976 | 5.775 | 825 | - |
| 915 | 5.15 | 1950 | 2.3 |
| 970 | 5.225 | 2550 | 3 |
| 973 | 5.625 | 1025 | 1.3 |
| 985 | 5.2 | 3000 | 3.6 |

Another three types of fiber with different core and cladding size are examined, the results of which are listed in Tables III-IV. We can obtain that the increment of threshold power by detuning pump wavelength is weakened with the increase of core-to-cladding ratio. This is due to that the increased core-to-cladding ratio will lead to higher upper state populations. This has the same effect as reducing the pump absorption coefficient, which tends to reduce hole burning and thus decrease MI threshold.

TABLE III
MI IN PM 26/400 FIBER

| Pump wavelength (nm) | Coupling frequency (kHz) | Threshold power (W) | |
|---|---|---|---|
| 976 | 3.975 | 600 | - |
| 915 | 3.4875 | 1150 | 1.9 |
| 970 | 3.5625 | 1500 | 2.5 |
| 973 | 3.8875 | 710 | 1.2 |
| 985 | 3.5625 | 1750 | 2.9 |

TABLE IV
MI IN PM 30/400 FIBER

| Pump wavelength (nm) | Coupling frequency (kHz) | Threshold power (W) | |
|---|---|---|---|
| 976 | 3 | 510 | - |
| 915 | 2.5375 | 815 | 1.6 |
| 970 | 2.6 | 1040 | 2 |
| 973 | 2.9125 | 565 | 1.1 |
| 985 | 2.575 | 1220 | 2.4 |

IV. EXPERIMENTAL STUDY

In this section, we presented the experimental study of pump wavelength on MI. To compare with the theoretical model, which considers mode coupling in one polarization direction, polarization-maintained (PM) fiber amplifier platform was employed. A monolithic, all-fiber, Yb-doped PM fiber amplifier is shown in Fig. 2. The seed laser in the experiment is a 50 mW linearly-polarized continuous-wave laser with central wavelength at 1064nm, which was then amplified in two pre-amplifier stages to ~25W. The main amplifier consists of a PM double-clad LMA Yb-doped fiber (YDF) with an Yb-doped core diameter of 21μm or 26μm and an inner clad diameter of 400μm, the core numerical aperture (NA) of which is about 0.064. The length of the gain fiber was chosen to maintain good amplifier efficiency, and the gain fiber is coiled loosely to rule out the influence of bending, which can also suppress MI [19]. A passive fiber is spliced to the gain fiber for power delivery, the output end of which is angle cleaved.

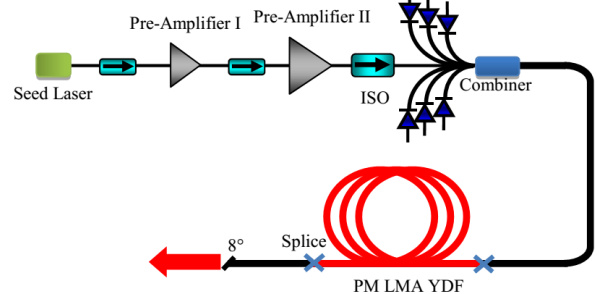

Fig. 2. The architecture of the all-fiber PM amplifier

MI thresholds for two type of fiber were marked in Fig. 3. The onset of MI is monitored by a photodetector with pinhole [8]. For 21/400 PLMA YDF, the amplifier pumped at 915nm shows a increase of threshold by a factor of 2 as compared to the fiber amplifier pumped at 976nm, which agree with the theoretical results. The slope near the maximum output power goes to zero with power rollover observed in Fig. 4(a), which is correlated with increased cladding light and also indicates the onset of MI. As predicated in the theoretical calculation, the improvement is weakened for fibers with larger core-cladding ratio: pumping with 915nm LD yielded only an increase by a factor of 1.5 in 26/400 PLMA YDF. Compared the experimental results fiber with the theoretical results, we can see that the suppression of MI agrees with our theoretical calculation, which demonstrated that our theoretical model is reliable. In addition, we can also conclude that higher MI suppression can be achieved when pumped at 985nm.

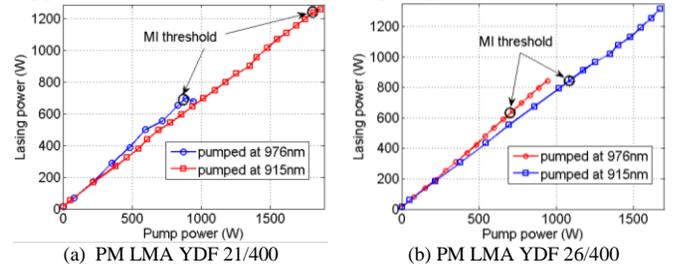

(a) PM LMA YDF 21/400         (b) PM LMA YDF 26/400
Fig. 3. Power characteristics of the all-fiber PM amplifier

The time trace of the in-pinhole power in the transition regimes was recorded, and the Fourier spectrum was shown in Fig. 4. One typical time trace is shown in Fig. 5(a), which exhibits oscillations with frequencies on the order of several kHz and indicates that the amplifier was operating in the transition regime. For 21/400 PM LMA YDF, the central frequency of resonances with certain bandwidth is around 4 kHz when pumped at 976nm. For 26/400 PM LMA YDF, the resonances showed up at around 2kHz when pumped at 976nm and reduced to ~1kHz when pumped at 915nm, which is due to that the reduced pump absorption coefficient lead to increase of gain saturation. Although the value of the experimental observed coupling frequency deviated from that predicated by theoretical model, which may be caused by pump power variation [20] and some temperature related parameters [21], the variation trend is the same: the maximal coupling frequency measured when pumped at 976nm is higher than that pumped at 915nm; the maximal coupling frequency is higher for smaller core size.

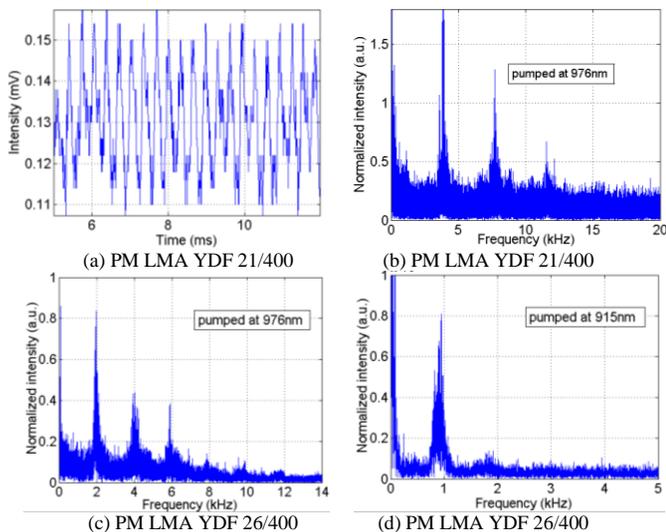

Fig. 4. Temporal and frequency characteristics of the all-fiber PM amplifier

## V. Conclusions

In summary, we have studied the influence of pump wavelength on MI theoretically and experimentally, which reveals that MI can be suppressed by simply detuning the wavelength of the pump light from the absorption peak at 976 nm. MI threshold power can be increased significantly by detuning the pump wavelength. By simply pumping at other wavelengths, such as 915nm, the threshold can be increased by a factor of ~2. Higher MI suppression can be achieved by detuning to 985nm. For amplifier employing gain fiber with larger core-to-cladding ratio, the enhancement of the threshold power by detuning the pump wavelength is smaller. The experimental results agree with theoretical calculation, which means that the model can be used to study MI and provide useful guidelines to suppress MI.